\documentclass[12pt]{article}
\usepackage[cp866]{inputenc}
\newcommand{\const}{{\rm\, const}}
\newcommand{\arctg}{{\rm\, arctg}}

\makeatletter
\renewcommand{\@oddhead}{October 2, 2008 \qquad\qquad\qquad\qquad\qquad\qquad Alexander Shatskiy, I.D. Novikov, N.S. Kardashev}

\input epsf

\begin{document}

\title{New analytic models of "traversable"\, wormholes}

\maketitle

\author{$\qquad\qquad\qquad\qquad$ Alexander Shatskiy${}^{1}$,
I.D. Novikov${}^{1,2,3}$, N.S. Kardashev${}^{1}$\thanks{
$\quad$\\$\quad$\\
${}^{1}$Astro Space Center, Lebedev Physical Institute, Moscow,
Russia, shatskiy@asc.rssi.ru \\
${}^{2}$Niels Bohr Institute, Blegdamsvej 17, DK-2100 Copenhagen,
Denmark \\
${}^{3}$Kurchatov Institute, Moscow, Russia}}

\begin{abstract}
The analytic solution of the general relativity equations for
spherically symmetric wormholes are given. We investigate the
special case of a "traversable"\, wormhole i.e., one allowing the
signal to pass through it. The energy-momentum tensor of wormhole
matter is represented as a superposition of a spherically
symmetric magnetic field and dust matter with negative matter
density. The dynamics of the model are investigated. We discuss
both the solution of the equation with a $\Lambda$-term and
without it. Superposing enough dust matter, a magnetic field, and
a $\Lambda$-term can produce a static solution, which turns out to
be a spherical Multiverse model with an infinite number of
wormholes connected spherical universes. Corresponding solution
can be static and dynamic.
\end{abstract}

\section{Introduction}
\label{s1}

A wormhole (WH) \cite{N-7}-\cite{N-24}) is a hypothetical object
described by a nonsingular solution of the Einstein equations with
two large (or infinite) space-time regions connected by a throat.
The two large space-time regions can be located in one universe or
belong to different universes in the Multiverse model (see
\cite{N-23}). In the last case, "traversable"\, WHs afford a
unique opportunity to explore other universes.

In the present paper, we analytically study the dynamics of a
spherical model of a non-equilibrium WH filled with matter. This
matter consists of a magnetic field and dust with negative mass
density. The obtained solution therefore generalizes the Tolman
solution \cite{Tolman} for a model with a spherically symmetric
electromagnetic field. As we see below, this generalization leads
to essentially new and important solutions. We use the method of
calculations of physical quantities in the frame comoving with
dust.

As the initial model for a WH, we use a static model in which
gravitational accelerations are everywhere identically zero.
Hence, the effective masses of both WH mouths vanish, although the
geometry of three-dimensional space is certainly non-Euclidean.
Such a model is considered in \cite{N-24} and \cite{Picon}, where
all matter is represented by a gravitating scalar field. We change
the scalar field into a superposition of an electro- magnetic
field and dust matter with negative energy density, which turns
out to be a methodologically important development and
generalization of these models. This allows us to apply methods of
Tolman's problem to calculate the model (see \cite{Tolman},
\cite{Oppenheimer}, \cite{Saibal}); these methods were generalized
and further developed by Shatskiy (see \cite{Shatsk1} and
\cite{Shatsk2}).

To further generalize and develop this method, we introduce the
cosmological $\Lambda$-term into the model. This allows obtaining
a principally new solution (see Section \ref{s7}) for a static
spherical model of the Multiverse. This model includes a infinite
number of spherical worlds connected by throats. To our knowledge,
this is the first analytic model of this type.

In the obtained solution, the Multiverse can have its total energy
density positive everywhere in space. In addition, this solution
can be generalized by the same method to the case of a dynamic
model shifted from equilibrium by an excess (or shortage) of dust
or by the $\Lambda$-term.

Some methodological details see in~\cite{ufn2}

\section{Einstein equations}
\label{s2}

We use the Armendariz-Picon static spherically symmetric solution
of the Einstein equations~\cite{Picon} for the description of the
unperturb WH model\footnote{ We use units where ${c=1}$ and
${G=1}$.}:
\begin{equation}
ds^2=dt^2 -dR^2 - r^2 (d\theta^2 + \sin^2\theta\, d\varphi^2)\,
,\quad r^2(R)=q^2+R^2  \, . \label{2-1}
\end{equation}
The minimum allowed radius of this WH is ${r_{_0}=q}$ (the radius
of the throat\footnote{Here and below, all values at the throat
have a subscript "${_0}$".}).

Usually, this solution is related to the energy-momentum tensor of
a scalar field (see~\cite{N-24}, \cite{Picon}). In the present
paper, we suggest another interpretation of this solution related
to a different representation of the energy-momentum tensor. This
new interpretation of solution (\ref{2-1}) can correspond to the
energy-momentum tensor represented as
\begin{equation}
T^n_k=\left(
\begin{tabular}{c c c c}
$-\frac{q^2}{8\pi r^4}$ & 0 & 0 & 0 \\
0 & $+\frac{q^2}{8\pi r^4}$ & 0 & 0 \\
0 & 0 & $-\frac{q^2}{8\pi r^4}$ & 0 \\
0 & 0 & 0 & $-\frac{q^2}{8\pi r^4}$ \\
\end{tabular}
\right)=\left(
\begin{tabular}{c c c c}
$+\frac{q^2}{8\pi r^4}$ & 0 & 0 & 0 \\
0 & $+\frac{q^2}{8\pi r^4}$ & 0 & 0 \\
0 & 0 & $-\frac{q^2}{8\pi r^4}$ & 0 \\
0 & 0 & 0 & $-\frac{q^2}{8\pi r^4}$ \\
\end{tabular}
\right)+\left(
\begin{tabular}{c c c c}
$-\frac{q^2}{4\pi r^4}$ & 0 & 0 & 0 \\
0 & 0 & 0 & 0 \\
0 & 0 & 0 & 0 \\
0 & 0 & 0 & 0 \\
\end{tabular}
\right) \label{2-2}
\end{equation}
The first term in the right-hand side corresponds to the
energy-momentum tensor of a static magnetic field with the
effective magnetic charge $q$ and the second term describes dust
matter with the negative energy density
\begin{equation}
\varepsilon_d=-\frac{q^2}{4\pi r^4} \label{e_d}
\end{equation}
This possible representation of the tensor ${T^n_k}$ is different
from its equivalent representation as the energy-momentum tensor
of a scalar field adopted in previous papers
\cite{Picon}-\cite{Picon4}.

An excess (or shortage) of dust (relative to ${\varepsilon_d}$) is
not in equilibrium and initiates motion. Because dust layers are
relatively independent, it is possible to integrate the equations
of motion for dust in a way similar to the solution of Tolman's
problem~\cite{Tolman}. In essence, it is the same as Tolman's
problem in a centrally symmetric and static electric (or magnetic)
field for uncharged dust.

We seek the metric tensor of the solution in the
form\footnote{Because dust is pressureless, we can choose the
synchronous commoving frame (with the time metric component equal
to 1).}:
\begin{equation}
ds^2=dt^2 - e^\lambda dR^2 - r^2 (d\theta^2 + \sin^2\theta\,
d\varphi^2) , \label{2-3}
\end{equation}
where $r^2$ and $e^\lambda$ are functions of both $R$ and $t$.
Below, we justify this choice of the metric.

We consider the problem in the presence of a $\Lambda$-term. The
Einstein equations corresponding to metric (\ref{2-3}) can be
written as\footnote{See, e.g., \cite{Landau1} for the derivation
(problem 5 in \S 100).}:
\begin{eqnarray}
8\pi T_t^t=8\pi\varepsilon +q^2/r^4+\Lambda = -e^{-\lambda}\left(
2r r_{,_{RR}} + r^2_{,_R}- r r_{,_R}\lambda_{,_R} \right)/r^2 +
\left(r r_{,_t}\lambda_{,_t}+r^2_{,_t} \right)/r^2 +1/r^2\, ,
\label{En1}\\
8\pi T_R^R = q^2/r^4 +\Lambda = \left( 2r r_{,_{tt}} +
r^2_{,_t}\right)/r^2  - e^{-\lambda} r^2_{,_R}/r^2 + 1/r^2\, ,
\label{En2}\\
8\pi T^R_t=0=e^{-\lambda}\left( 2r_{,_{Rt}}-r_{,_R}\lambda_{,_t}
\right)/r \, ,
\label{En3}\\
8\pi T^\theta_\theta =8\pi T^\varphi_\varphi=-q^2/r^4 +\Lambda =
-\left\{e^{-\lambda}\left[2r_{,_{RR}} -
r_{,_R}\lambda_{,_R}\right]
-2r_{,_{tt}}-r\lambda_{,_{tt}}-r\lambda_{,_t}^2/2-r_{,_t}\lambda_{,_t}\right\}/(2r)
\, . \label{En4}
\end{eqnarray}
{\bf  Thus, the energy-momentum tensor includes three types of
matter:\\
a centrally symmetric magnetic field, \\
a cosmological $\Lambda$-term, and \\
dust matter with the density\footnote{In the absence of the L
term, solution (\ref{2-3}) coincides with (\ref{2-1}) only for the
density $\varepsilon$ equal to $\varepsilon_d$ [see
.~(\ref{e_d})].} $\varepsilon$.}

\section{Solution and the Initial Conditions}
\label{s3}

Equation (\ref{En3}) can be integrated with respect to time:
\begin{equation}
e^{-\lambda}r^2_{,_R}=F_1(R) \, . \label{3-1}
\end{equation}
At the throat, the condition ${r_{,_R}=0}$ must be satisfied, and
therefore
\begin{equation}
F_1(0)=0 \, . \label{3-1-2}
\end{equation}
By substituting (\ref{3-1}) in Eqn (\ref{En2}), we obtain
\begin{equation}
\frac{q^2}{r^2} +\Lambda r^2= \left( r r^2_{,_t}
\right)_{,t}/r_{,_t} -F_1+1 \, . \label{3-2}
\end{equation}
Integrating this equation with respect to time yields
\begin{equation}
\frac{q^2}{r}-\Lambda r^3/3+r r^2_{,_t}+(1-F_1) r = F_2(R) \, .
\label{3-3}
\end{equation}
The functions ${F_1(R)}$ and ${F_2(R)}$ determine the initial
conditions for the velocity and dust density distributions.

With ${\lambda_{,_t}}$ and ${\lambda_{,_{tt}}}$  expressed from
Eqn (\ref{En3}), Eqn (\ref{En4}) can be rewritten as
\begin{equation}
\left[ q^2/r^2+\Lambda r^2+e^{-\lambda}r_{,_R}^2 - r_{,_t}^2 - 2r
r_{,_{tt}}\right]_{,_R}=0 \, . \label{En4-2}
\end{equation}
When Eqn (\ref{En2}) is satisfied, integrating Eqn (13) yields an
identity. {\bf  This identity is the consequence of setting the
time component of metric (\ref{2-3}) to unity, which justifies
this choice of the metric.}

It is worth noting from the methodological standpoint that this
form of the metric for dust is obvious because dust is
pressureless and hence the comoving frame is simultaneously a
synchronous one. But this is not so obvious when a spherically
symmetric electromagnetic field is added, and the obtained
identity is a consequence of the absence of interaction between
uncharged dust and the electromagnetic field. Moreover, the
Lorentz transformation along the magnetic field lines does not
change this field, which justifies the use of the dust comoving
reference frame. For motion in this frame (along the field), the
electromagnetic field remains invariant. If the metric could not
be represented in form (\ref{2-3}), further calculations would be
impossible. It is this non-obvious fact that allowed us to
successfully develop our method and to obtain all important
results.

Multiplying Eqn (\ref{En1}) by ${r^2 r_{,_R}}$ and expressing
${\lambda_{,_t}}$ from Eqn (\ref{En3}), we obtain
\begin{equation}
8\pi\varepsilon r^2 r_{,_R} +\left[ \Lambda r^3/3-q^2/r
-r(1-F_1)-r r_{,_t}^2\right]_{,_R} =0\, . \label{En1-2}
\end{equation}
Integrating this with respect to $R$ from $0$ to $R$ with account
for (\ref{3-1-2}) yields
\begin{equation}
\int\limits_0^R 8\pi\varepsilon r^2 r_{,_R}\, dR -q^2/r +\Lambda
r^3/3 -r(1-F_1)-r r_{,_t}^2 =-q^2/r_{_0}+\Lambda r_{_0}^3/3-
r_{_0}- r_{_0} r_{_0,_t}^2\, . \label{3-6}
\end{equation}
Here, we took into account that the function ${r(R=0)= r_{_0}(t)}$
can be time-dependent.

Expressing ${r r_{,_t}^2}$ from (\ref{3-3}) and substituting it in
(\ref{3-6}), we obtain
\begin{equation}
\int\limits_0^R 8\pi\varepsilon r^2 r_{,_R}\, dR = F_2(R)
-q^2/r_{_0}-r_{_0}+\Lambda r_{_0}^3/3 -r_{_0} r_{_0,_t}^2 \, .
\label{3-7}
\end{equation}
The derivative of this expression with respect to $R$ yields one
more integral of motion (which expresses mass conservation in the
comoving volume):
\begin{equation}
8\pi\varepsilon r^2 r_{,_R}=\frac{d F_2}{dR} \, . \label{bianki1}
\end{equation}
The Bianchi identities\footnote{Here, $g$ is the determinant of
the metric tensor.}:
\begin{equation}
T^n_{k; n} =\frac{1}{\sqrt{-g}}\,\frac{\partial
(T^n_k\sqrt{-g})}{\partial x^n} - \frac{T^{nl}}{2}\,\frac{\partial
g_{nl}}{\partial x^k} =0 \label{bianki0}
\end{equation}
are contained in the Einstein equations, their components ${k=t,
\theta, \varphi}$ vanish identically, and the component ${k=R}$
yields the result in (\ref{bianki1}), which is an analogue of the
integral for the pure-dust solution (see~\cite{Landau1}, $\S$103)
in the theory of the evolution of a dust cloud. The explicit form
of the function ${e^\lambda (R,t)=r_{,_R}^2/F_1}$ and an implicit
form of the function ${r^2(R,t)}$ are obtained in
Section~\ref{s5}.

%\section{Initial conditions}
\label{s4}

{\bf Below, we use the index "i"\, to denote all quantities at the
initial instant ${t=0}$.}

The velocity distribution is set to zero initially:
\begin{equation}
\left. r_{,_t}^2 \right|_{(t=0)}=0 \label{ini1}
\end{equation}
A nonzero initial velocity is considered in Appendix~\ref{ap1}. We
note that in all cases, the choice of initial conditions is not
arbitrary and must satisfy constraints imposed by (\ref{En1}) and
(\ref{En3}). These equations do not contain second time
derivatives and, as is well known (see \cite{Picon3}), under the
condition that ${t=\const}$ at the initial cross section, they are
automatically satisfied for all values $t$ by virtue of the other
Einstein equations.

We choose the initial coordinate scale of $R$ in (\ref{2-3}) such
that
\begin{equation}
r_i = \sqrt{q^2+R^2}\, ,\qquad
\varepsilon_{di}\equiv\varepsilon_{d}(r_i) = -\frac{q^2}{4\pi
r_i^4}\, . \label{ini2}
\end{equation}
where ${r_i(R)}$ coincides with (\ref{2-1}) and Eqn (\ref{3-1-2})
is automatically satisfied. We set
\begin{equation}
s(r_i)\equiv\frac{1}{r_i}\int\limits_0^R
8\pi(\varepsilon_i-\varepsilon_{di})r_i^2 r_{i,_R}\, dR
 =\frac{1}{r_i}\int\limits_q^{r_i} 8\pi\varepsilon_i r_i^2 \, dr_i +
2\frac{q}{r_i}\left(1-\frac{q}{r_i}\right) \label{3-7-1}
\end{equation}
The quantity $s$ has the meaning of the doubled Newtonian
potential of dust matter excessive relative to $\varepsilon_d$
[see~(\ref{ini2})]. The relation of this potential to the
functions $F_1$ and $F_2$ follows from (\ref{3-6}) and (\ref{3-7})
at the initial instant:
\begin{equation}
F_1= 1-s-\frac{q^2}{r_i^2}-\frac{\Lambda}{3}\left( r_i^2-q^3/r_i
\right)\, , \label{3-8}
\end{equation}
\begin{equation}
F_2= r_i\left(s+2\frac{q^2}{r_i^2}-\frac{\Lambda q^3}{3r_i}\right)
\, . \label{3-9}
\end{equation}
Equation (\ref{3-1}) implies that ${F_1\ge 0}$. According to
(\ref{3-8}), this imposes constraints on the distribution
${s(R)}$.

From Eqn (\ref{En2}), we can find ${r_{,_{tt}}}$ at the initial
instant:
\begin{equation}
\makebox{at }t=0 \, :\quad  r_{,_{tt}}=
-\frac{s}{2r_i}+\frac{\Lambda r_i}{3}+\frac{\Lambda q^3}{6 r_i^2}
\label{aks1}
\end{equation}
Equation (\ref{aks1}) shows that for ${\Lambda =0}$, the dynamics
are absent in the region were ${s=0}$ (at least up to the instant
of intersection of the dust layers, when the model considered
becomes inapplicable).

\section{Solution without the $\Lambda$-term}
\label{s5}

We consider the case ${\Lambda =0}$ in more detail.

For ${\Lambda =0}$, it follows from Eqn (\ref{aks1}) that {\bf the
WH throat size remains constant ${[r_{_0}(t)=q]}$ in the whole
region of the allowed values of $R$ and $t$ for the sought
solution, until matter starts flowing through the throat. In
addition, expression (\ref{aks1}) implies that the dynamics of
matter in a WH depend only on the internal layers of matter (with
smaller values of r) and are independent of the external layers.
Therefore, matter on one side of the throat does not influence
matter on the opposite side (unless there is matter flowing
through the throat or layer intersection occurs). }

It can be verified directly from (\ref{aks1}) that at the initial
instant, the acceleration ${r_{,_{tt}}}$ is negative for ${s>0}$.
Hence, the excessive mass of dust is to collapse.

We introduce the definition of the apparent horizon.

\begin{figure}[t]
\centering \epsfbox[20 20 440 420]{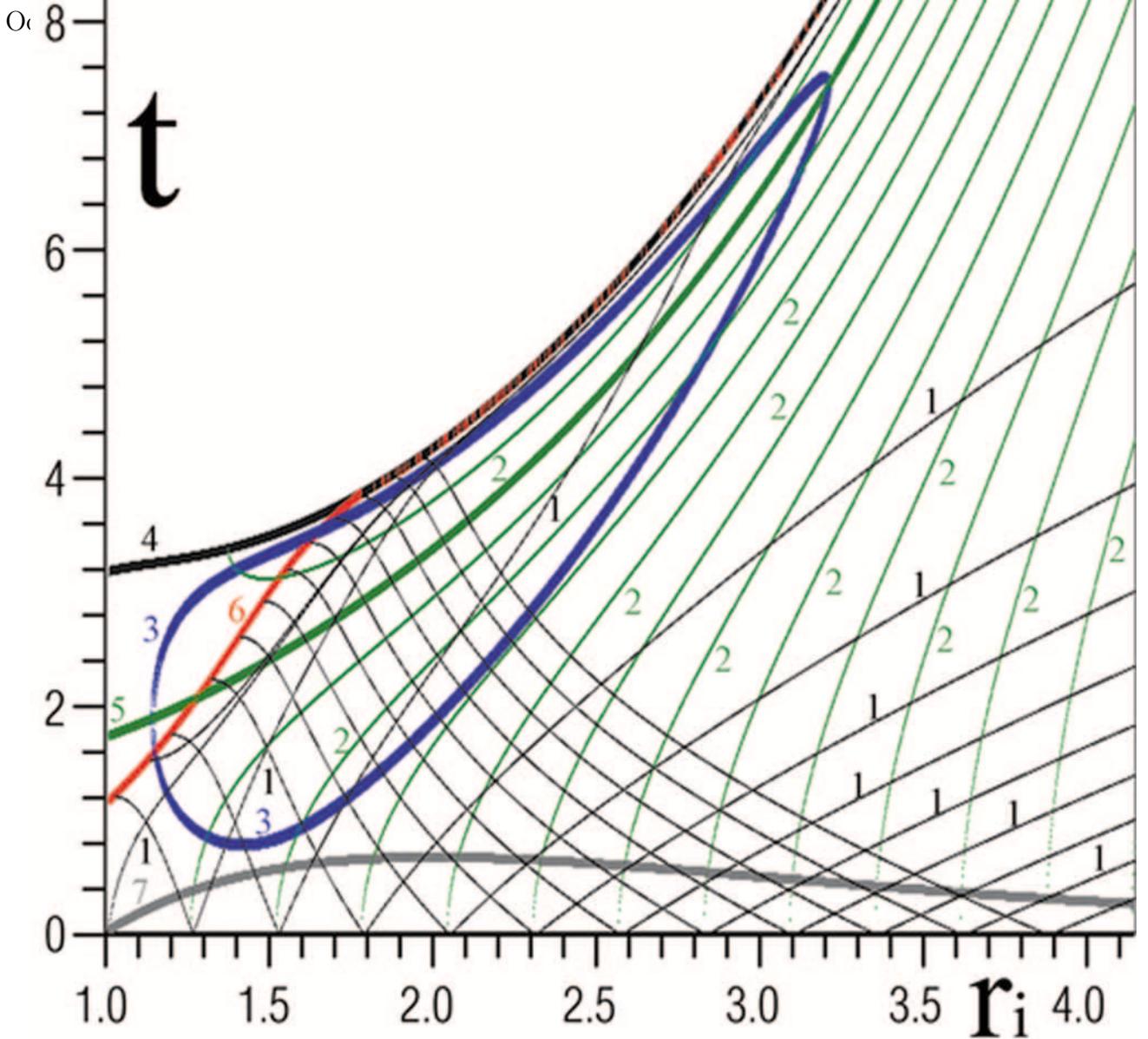} \caption{Diagrams
${t(r_i/q)}$: 1~-~for light geodesics; 2~-~for ${r=\const}$;
3~-~for the apparent horizon and the inner horizon; 4~-~for the
limiting stopping time; 5~-~for the throat ${r=q}$; 6~-~for the
intersection times of adjacent dust layers; 7~-~the dependence
${s(r_i/q)= 0.7(r_i/q-1)\cdot\exp(2-r_i/q)}$ --- the scale for $s$
(vertical axis) coincides with the time scale $t$.} \label{R1}
\end{figure}

{\bf  The criterion of the absence of a apparent horizon during
the collapse is given by}
\begin{equation}
V<1 \label{V}
\end{equation}
for the dust velocity [see also
\cite{Shatsk1},\cite{Nov1},\cite{Nov2},\cite{Shatsk2} and
(\ref{3-3})], where the dust velocity $V$ is determined as
\begin{equation}
V^2\equiv r_{,_t}^2 e^\lambda /r_{,_R}^2 = r_{,_t}^2/F_1 =
\frac{(s+2q^2/r_i^2)r_i /r -s-q^2/r_i^2 - q^2/r^2}{1-s-q^2/r_i^2}
\label{V2}
\end{equation}
For ${q=0}$, this result coincides with the solution for
collapsing dust (see~\cite{Shatsk1}).

To avoid a ${0/0}$ ambiguity for the function $V^2$ at the throat
at the initial instant ${(F_1[r_i\to r_{_0}]\to 0)}$, it is
necessary to ensure a more rapid convergence to zero of the
function ${r_{,_t}^2 (t=0, r_i\to r_{_0})}$ than the function
${F_1(r_i\to r_{_0})}$ (see Appendix~\ref{ap1}).

\begin{figure}[t]
\centering \epsfbox[20 20 440 410]{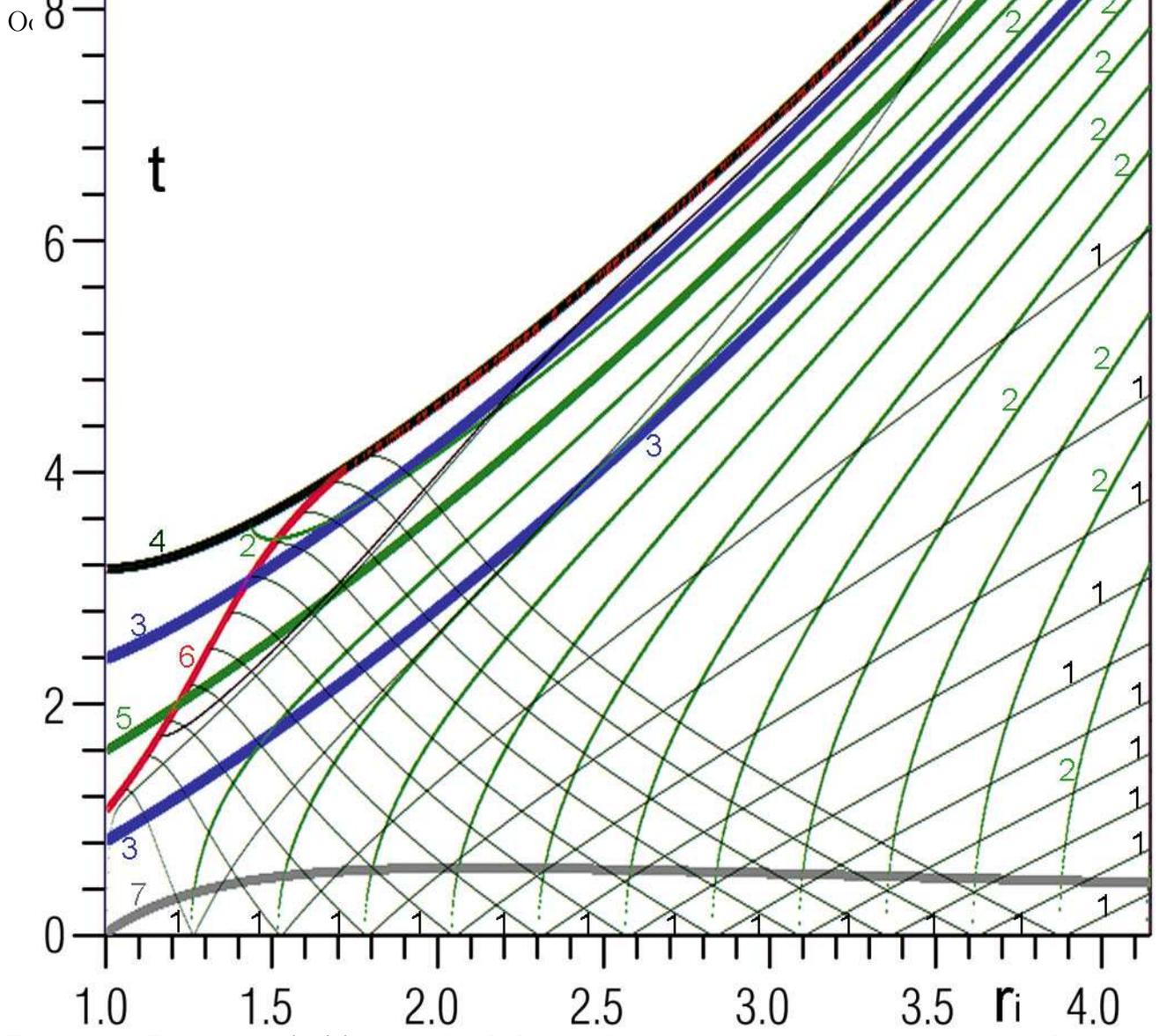} \caption{Diagrams
${t(r_i/q)}$: 1~-~for light geodesics; 2~-~for ${r=\const}$;
3~-~for the apparent horizon and the inner horizon; 4~-~for the
limiting stopping time; 5~-~for the throat ${r=q}$; 6~-~for the
intersection times of adjacent dust layers; 7~-~the dependence
${s(r_i/q)=(1-q/r_i)(2+0.5(1-q/r_i))q/r_i}$ --- the scale for $s$
(vertical axis) coincides with the time scale $t$.} \label{R2}
\end{figure}

The law of motion of a dust layer can be found from (\ref{3-3}):
\begin{equation}
t=\int\limits_{r_i}^r\frac{-r\, dr}{\sqrt{(F_1-1)r^2+F_2r-q^2}}=
\int\limits_{r_i}^r\frac{-r\, dr}{r_i \sqrt{(1-r/r_i)
[(s+q^2/r_i^2)r/r_i-q^2/r_i^2]}} \label{3-4}
\end{equation}
This quadrature can be elementarily integrated:
\begin{equation}
t=r_i
\frac{\sqrt{(1-r/r_i)[(s+q^2/r_i^2)r/r_i-q^2/r_i^2]}}{s+q^2/r_i^2}+
r_i \frac{s+2q^2/r_i^2}{2(s+q^2/r_i^2)^{3/2}}\cdot \arccos \left[
1-\frac{2(s+q^2/r_i^2)(1-r/r_i)}{s}\right] \label{3-10-alt}
\end{equation}
or in another form,
\begin{equation}
t=r_i\left[
\frac{\sqrt{(1-r/r_i)[(s+q^2/r_i^2)r/r_i-q^2/r_i^2]}}{s+q^2/r_i^2}+
\frac{s+2q^2/r_i^2}{(s+q^2/r_i^2)^{3/2}}\cdot
\arctg\sqrt{\frac{(s+q^2/r_i^2)(1-r/r_i)}{(s+q^2/r_i^2)r/r_i-q^2/r_i^2}}
\right] \label{3-10}
\end{equation}
{\bf Hence, the motion of dust layers depends on the excessive
potential $s$ and the initial distance from the WH throat:
$r_i/q$.}

However, this model is physically correct only until a possible
intersection of dust layers. This is because after the
intersection, an incoming energy flux appears in the
matter-comoving frame (from the intersected matter layers), which
has not been taken into account in the energy-momentum tensor. The
intersection of adjacent matter layers corresponds to an infinite
energy density. This corresponds to the condition ${r_{,_R}=0}$
(or ${\frac{dr}{dr_i}=0}$) [see~(\ref{ini2})]. Differentiation of
(\ref{3-10}) with respect to ${r_i}$ allows explicitly determining
the function ${\frac{dr}{dr_i}(r,s,q)}$. Differentiation of the
left-hand side of (\ref{3-10}) yields zero. We let a prime denote
the derivative with respect to $r_i$, express all distances in the
units of units $r_i$, and omit intermediate bulky calculations;
the final expression for this function is given by
\begin{eqnarray}
r'=1+\frac{(1-r)\left[ s(s'-2q^2)(2q^2-r\{s+q^2\})-
q^2(s+2q^2)(2s'+s) \right]} {2rs (s+q^2)^2}+
\nonumber \\
+\frac{\left[ 2s (s+q^2)+(2q^2-s')(s+4q^2) \right]
\sqrt{(1-r)[(s+q^2)r-q^2]}} {2r (s+q^2)^{5/2}}\cdot
\arctg\sqrt{\frac{(s+q^2)(1-r)}{(s+q^2)r-q^2}} \label{rs}
\end{eqnarray}
It is impossible to use this formula to determine the instant of
intersection of nonadjacent layers because the intersection with
nonadjacent layers does not lead to any irregularities for dust.
Clearly, the intersection of nonadjacent layers occurs after the
intersection of adjacent ones and cannot propagate with a
superluminal velocity. Therefore, after constructing diagrams
$t(r_i)$ for light cones and the curve corresponding to the
intersection of adjacent layers ${(r'=0)}$, we can determine
regions that are definitely free from layer intersection (see the
Figure~\ref{R1} or \ref{R2}).

Using formula (\ref{V2}), we can reach several important
conclusions.

{\bf  1. The formation of a horizon is possible only for a
sufficiently large parameter $s$:}
\begin{equation}
s>2\frac{q}{r_i}\left(1-\frac{q}{r_i}\right) \label{s-hor}
\end{equation}
{\bf After the appearance of a horizon, the WH becomes
"non-traversable".} However, dust layers can already start
intersecting before the horizon is reached.

{\bf  2. Values of the function ${t(r,r_i)}$ are bounded by the
maximum time $t_{stop}$ (from the beginning of motion until the
stop).} In (\ref{3-10-alt}), this time corresponds to the argument
of ${\bf \arccos}$ being minus unity (or to the zero value of the
square root expression in the same formula):
\begin{equation}
t_{stop}=\frac{\pi r_i(s+2q^2/r_i^2)}{2(s+q^2/r_i^2)^{3/2}}
\label{3-10-2}
\end{equation}
The time $t_{stop}$ corresponds to the deviation ${\Delta r\approx
sr_i^3/q^2}$ from the initial position. {\bf  For small deviations
(corresponding to small values of $s$), harmonic oscillations with
the period ${T=2t_{stop}\approx 2\pi r_i^2/q}$ occur} (see
Appendix~\ref{ap3}).

{\bf  3. The existence of the second (smaller) root of the
equation ${V^2(r)=1}$ implies the appearance of the second (inner)
horizon in the system (see the Figures~\ref{R1} and \ref{R2}).}

4. For parameter ${s=2q(1-q/r_i)/r_i}$ (or ${\varepsilon =0}$) the
scalar ${\bf g^{ik}\, r_{;i} r_{;k}=(1-q/r)^2}$ (scalar equal to
the scalar of the extremal Reisner-Nordstrem solution). Therefore:
{\bf if the energy density tends to zero --- the solution
(\ref{3-10}) tends to the extremal Reisner-Nordstrem solution of
black hole (in the comoving, free-falling frame).}

\section{Solution for $s=0$ with the $\Lambda$-term}
\label{s6}

We consider the case where $s=0$. Eqn (\ref{En2}) with (\ref{3-1})
and (\ref{3-8}) then becomes
\begin{equation}
2r r_{,_{tt}} = \Lambda r^2 +q^2\left(\frac{1}{r^2}
-\frac{1}{r_i^2}\right)- \frac{\Lambda}{3}\left( r_i^2-q^3/r_i
\right) - r^2_{,_t} \, , \label{5-1}
\end{equation}
and Eqn (\ref{3-3}) takes the form
\begin{equation}
r^2_{,_t} = (r-r_i) \left[ -\frac{q^2}{r^2 r_i^2}(r-r_i)
+\frac{\Lambda}{3} \left( r+r_i+\frac{q^3}{rr_i} \right) \right]\,
. \label{5-2}
\end{equation}
These two equations imply that at $s=0$, matter starts expanding
from the rest state (see~\ref{ini1}): inflation due to the
$\Lambda$-term). {\bf If the dimensionless parameter
\begin{equation}
a \equiv \Lambda q^2 \label{5-3}
\end{equation}
exceeds some critical value ${a_{cr}}$, the inflation continues
unlimitedly in the entire volume ${(q\le r_i\le r<\infty)}$ until
an outer horizon forms, ${V^2(r)=1}$.} The value ${a_{cr}}$ is
found from the condition of the maximum of the expression for the
parameter $a$, which is obtained by equating the second factor (in
square brackets) in Eqn (\ref{5-2}) to zero:
\begin{equation}
a_{cr} =MAX \left[ \frac{3\tilde q^4 (1-\tilde r) \tilde r^2}
{1+\tilde r+\tilde r^2 \tilde q^3} \right] \approx  0.22 \, ,\quad
\tilde r\equiv r_i/r\, ,\quad  \tilde q\equiv q/r_i \, .
\label{5-3}
\end{equation}
{\bf  For ${a<a_{cr}}$, a region appears where the inflation
stops.} This region starts emerging at the throat and extends
toward larger values of $r_i$ as the parameter $a$ decreases.

{\bf  As ${r/r_i\to\infty}$, the inflation occurs exponentially:
${r\propto\exp(t\sqrt{\Lambda /3})}$.}

The function $r'$ can be expressed in quadratures (see
Appendix~\ref{ap2}):
\begin{eqnarray}
r' = 1+\sqrt{r-r_i}\exp\left(-\int\limits_{r_i}^r P_2(x)\,
dx\right) \cdot \int\limits_{r_i}^r
\frac{Q_2(r)-P_2(r)}{\sqrt{r-r_i}} \exp\left(\int\limits_{r_i}^r
P_2(x)\, dx\right) \, dr \, ,
\nonumber \\
P_2=\frac{\frac{r_i^2}{2r^2}-\frac{r_i^2(r-r_i)}{r^3}-\frac{a}{2}
\left(1-\frac{q^3}{r^2 r_i} \right)}
{-\frac{r_i^2(r-r_i)}{r^2}+a\left( r+r_i+\frac{q^3}{r r_i}
\right)} \, ,\quad Q_2-P_2=\frac{a+(r+r_i)\left(
\frac{r_i(r-r_i)}{r^3}-\frac{aq^3}{2r^2 r_i^2} \right)}
{-\frac{r_i^2(r-r_i)}{r^2}+a\left( r+r_i+\frac{q^3}{r r_i}
\right)} \, . \label{rs2}
\end{eqnarray}
Using this equation, we can find regions where the inflation
occurs without dust layer intersections.

\section{Model of the Multiverse}
\label{s7}

Dust with the positive $s$ initially accelerates toward the center
(without the $\Lambda$-term). The $\Lambda$-term without excessive
dust leads to the original inflationary solution
[see~(\ref{aks1})]. At the WH throat, the potential $s=0$, and the
$\Lambda$-term at the throat provides nonzero matter acceleration.
This contribution cannot be compensated at or near the throat
[see~(\ref{En2})], where ${r_{,_t}=r_{,_{tt}}=0}$ and ${r=q}$.
Hence, there is no static solution for a WH with the throat radius
${r_{_0}=q}$ and the $\Lambda$-term.

Nevertheless, a static solution exists for a WH with the
$\Lambda$-term and ${r_{_0}\ne q}$. This solution can be easily
derived from Eqns (\ref{En1}), (\ref{En2}) and (\ref{En4})  (with
${r_{,_t}=0}$ and ${r_{,_{tt}}=0}$).

The metric of a static WH is determined by expression (\ref{2-1}),
taking into account that ${r^2(R)\ne q^2+R^2}$.

From (6), we obtain ${r_{,_R}^2=1-q^2/r^2-\Lambda r^2}$ and easily
find the expression for the throat radius ${r_{_0}}$,
\begin{equation}
\Lambda r_{_0}^4 -r_{_0}^2+q^2=0 \quad\Rightarrow\quad
r_{_0}^2=\frac{1-\sqrt{1-4\Lambda q^2}}{2\Lambda}\to q^2(1+\Lambda
q^2) \qquad(\makebox{at  }\Lambda q^2\to 0) \, , \label{6-1}
\end{equation}
and the dependence ${r(R)}$:
\begin{equation}
r^2(R)=\frac{1-\sqrt{1-4\Lambda q^2}\cdot\cos
(2\sqrt{\Lambda}R)}{2\Lambda} \label{6-3}
\end{equation}
The distribution of $\varepsilon$ for this solution is
determined\footnote{Here, the unknown value
${r_{,_{RR}}=q^2/r^3-\Lambda r}$ can be obtained from expression
(\ref{En4}) or by directly differentiating Eqn (\ref{6-3}).} from
(\ref{En1}):
\begin{equation}
\varepsilon = \frac{\Lambda}{4\pi}-\frac{q^2}{4\pi r^4}
\label{6-2}
\end{equation}
Thus, for the static solution with the $\Lambda$-term, the throat
radius and dust density are larger than without the
$\Lambda$-term.

The total energy density for the static solution with the
$\Lambda$-term is
\begin{equation}
T^t_t =\frac{3\Lambda}{8\pi}-\frac{q^2}{8\pi r^4} \label{Ttt}
\end{equation}
The condition for the value ${T^t_t}$ to be nonnegative everywhere
in space (for ${r\ge r_{_0}}$) has the form
\begin{equation}
T^t_t \ge 0 \quad \makebox{at } a=\Lambda q^2 \ge \frac{3}{16}
\label{Ttt2}
\end{equation}
{\bf Therefore, in the presence of the $\Lambda$-term, solutions
for wormholes with a positive total energy density can be found.}

From (\ref{6-3}), we can obtain the maximum allowed radius of the
static metric with the $\Lambda$-term:
\begin{equation}
r_{max}^2=\frac{1+\sqrt{1-4\Lambda
q^2}}{2\Lambda}\to\frac{1}{\Lambda} \qquad (\makebox{at  }\Lambda
q^2\to 0) \, . \label{6-4}
\end{equation}
Beyond this radius, the Universe starts contracting again until a
new throat occurs.

{\bf  Solution (\ref{6-3}) describes a static Multiverse with an
infinite number of throats. When there is no charge ${(q=0)}$,
this solution becomes a cosmological solution without wormholes,
corresponding to a closed isotropic universe} (see~\S 112,
\cite{Landau1}).

When there is an excess (or shortage) of the dust part of the
energy density or the $\Lambda$-term relative to (\ref{6-2}), the
obtained solution for the Multiverse becomes dynamical. Its
analytic study is complicated by the need to solve a fourth-order
algebraic equation and to calculate a quadrature similar to
(\ref{rs2}) for the instant of dust layer intersection.

\section{Conclusion}
\label{s12}

In this paper, we have generalized and developed the method
suggested earlier in papers \cite{Tolman}, \cite{Saibal} and
applied it to new problems in modern cosmology. We found and
analyzed analytic solutions of the general relativity equations
describing the dynamics of a traversable wormhole. The results
obtained are important for the analysis of the general properties
of traversable wormholes. We also obtained a solution describing a
spherically symmetric model of the Multiverse. We have not
analyzed the properties of geodesics that describe the motion of
particles and other matter (energy and information) through a
wormhole and its vicinity. These problems, which are mostly
important for the analysis of the possible observational
appearance of such objects, are considered elsewhere (see, e.g.,
\cite{N-17}).

To conclude, we remark on the term "Multiverse". This term is used
in physics and cosmology in two different senses.

First, this term assumes the possibility of the parallel co-
existence of many or even an infinite number of different worlds,
possibly emerged from a quantum vacuum (in some sense, in
different places at different times). In this paper, this term is
used exactly in this sense.

Second, this term is sometimes used to denote a sample of
different realities in the Everett interpretation of quantum
mechanics. The usage of one term for different notions is
sometimes confusing. In our opinion, different terms should be
used for these concepts. We propose keeping the term
"Multiverse"\, for cosmology, and refer to the set of Everett
worlds (following the proposal by M.B. Mensky) as the
"Alterverse", keeping in mind different classical alternatives of
the Everett world.

\section{Acknowledgments}
\label{s11}

We are grateful to the staff of the theoretical astrophysics
department of the Lebedev Physical Institute for the discussions.
The work is supported by the RFBR grants ${07-02-01128a}$ and
${08-02-00090a}$, grants of scientific schools ${NSh-626.2008.2}$
and ${NSh-2469.2008.2}$, and the program of the Russian Academy of
Sciences, "The Origin and Evolution of Stars and Galaxies 2008".

$\quad $\\
\hrule

\appendix
\section{Initial conditions with a nonzero velocity}
\label{ap1}

At the WH throat, the expression for $V^2$ in (\ref{V2}) involves
a ${0/0}$ ambiguity. To resolve this ambiguity, we consider the
model with a nonzero initial velocity of matter. We introduce the
notation
\begin{equation}
1-q/r_i\equiv\alpha \, ,\quad r_i/r-1\equiv \beta \, ,\quad \left.
r_{,_t}^2 \right|_{(t=0)} \equiv \gamma\, . \label{p1-1}
\end{equation}
and assume that ${\Lambda =0}$. The expressions for the functions
$F_1$ and $F_2$ with ${\gamma\ne 0}$ [see~(\ref{3-6}) and
(\ref{3-3})] are given by
\begin{equation}
F_1=1-s -q^2/r_i^2+\gamma -\gamma_{_0} q/r_i \, ,\quad
F_2=r_i\left[ s+2q^2/r_i^2+\gamma_{_0}q/r_i \right] \, .
\label{p1-3}
\end{equation}
With (\ref{3-3}) and a nonzero initial velocity, the expression
for $V^2$ in (\ref{V2}) takes the form
\begin{equation}
V^2=
\frac{-s-q^2/r_i^2+\gamma-\gamma_{_0}q/r_i+(s+2q^2/r_i^2+\gamma_{_0}q/r_i)r_i/r
-q^2/r^2}{1-s -q^2/r_i^2+\gamma -\gamma_{_0} q/r_i } \label{V2-2}
\end{equation}
In the vicinity of the WH throat close to the initial instant, the
functions $\alpha$ and $\beta$ take small values. From
(\ref{3-7-1}), in the linear order in $\alpha$, we have
$$
s\approx \alpha \kappa\, ,\quad \kappa=8\pi
r_i^2(\varepsilon_i-\varepsilon_d)\approx
8\pi\varepsilon_{_0}q^2+2 <2 \, .
$$
We consider Eqn (\ref{V2-2}) in the linear order in $\alpha$ and
$\beta$. After all transformations, we finally obtain
\begin{equation}
V^2\approx \frac{\gamma +\gamma_{_0}\beta} {\gamma
-\gamma_{_0}(1-\alpha)+\alpha(2-\kappa)} \to
\frac{\gamma_{_0}(1+\beta)} {\alpha(\gamma_{_0}+2-\kappa)}
\makebox{   at }\gamma\to\gamma_{_0}\, . \label{V2-3}
\end{equation}
It follows that the condition for $V^2$ to be nonsingular at the
throat is that ${\gamma_{_0}=0}$. Then
\begin{equation}
V^2\approx \frac{\gamma} {\gamma +\alpha(2-\kappa)} \label{V2-4}
\end{equation}
Expression (\ref{V2-4}) for $V^2$ is regular and has no
ambiguities if the function $\gamma$ tends to zero at the throat
faster than the function $\alpha$. In this case, the rate of
change of the radius $r$ vanishes at the throat (as must be the
case by the definition of the throat). In the linear order in
$\beta$, the function $V^2$ is time-independent near the throat.

\section{Study of intersections of adjacent layers}
\label{ap3}

We consider the model without the $\Lambda$-term for $s>0$. In
this case, the excessive matter collapses, i.e., ${r\le r_i}$.

{\bf Lemma. There always exists a nonzero time interval
${[0,t]}$ inside which no intersections of adjacent layers occur.} \\
Equation (\ref{rs}) implies that this is obvious for a nonzero
potential ${s}$. The only point at which this statement must be
proved is the point corresponding to ${s=0}$.

We study the asymptotic regime ${s\to 0}$ in more detail:
\begin{equation}
s<<q^2/r_i^2\le 1 \label{p3-1}
\end{equation}
From the allowed range of the radius (until the stopping point)
${r \ge q^2/(sr_i+q^2/r_i)}$, we deduce
\begin{equation}
1-\frac{r}{r_i}\le \frac{s}{s+q^2/r_i^2}\le \frac{s}{q^2/r_i^2}<<1
\label{p3-1-2}
\end{equation}
Because the last term in (\ref{rs}) contains no $s$ in the
denominator, it can be neglected compared to the previous term.
This can be proved in more detail by considering the there
exhaustive options for the ratio of the numerator and denominator
in the square root expression of the ${\bf \arctan}$:

1. ${(s+q^2/r_i^2)(1-r/r_i) << (s+q^2/r_i^2)r/r_i-q^2/r_i^2}$

2. ${(s+q^2/r_i^2)(1-r/r_i) \sim (s+q^2/r_i^2)r/r_i-q^2/r_i^2}$

3. ${(s+q^2/r_i^2)(1-r/r_i) >> (s+q^2/r_i^2)r/r_i-q^2/r_i^2}$

Keeping the leading terms, we obtain the asymptotic form of Eqn
(\ref{rs}):
\begin{equation}
r'\to 1+\frac{(1-r/r_i)[-q^2\cdot 2q^2\cdot 2r_i s']}{2s
q^4}=1-2r_i s' (1-r/r_i)/s \label{p3-2}
\end{equation}
In a similar way, the first term in the right-hand side of
expression (\ref{3-10-alt}) can be neglected (compared to the
second term), and hence asymptotically with respect to time, we
have
\begin{equation}
\cos\left(\frac{tq}{r_i^2}\right)\to 1-\frac{2q^2}{s
r_i^2}(1-r/r_i) \quad\Longrightarrow\quad r \to r_i + \frac{s
r_i^3}{2 q^2}\left[\cos\left(\frac{tq}{r_i^2}\right)-1 \right]
\label{p3-3}
\end{equation}
Comparing Eqns (\ref{p3-2}) and (\ref{p3-3}), we obtain the sought
asymptotic form as ${s\to 0}$:
\begin{equation}
r'\to 1-\frac{2r_i^3 s'}{q^2}\sin^2\left(\frac{t q}{2r_i^2}
\right) . \label{p3-4}
\end{equation}
This proves the lemma. \\
In addition, Eqn (\ref{p3-3}) implies harmonic dynamics
(oscillations) for ${s\to 0}$ and ${2r_i^3 s'/q^2<1}$.

\section{Quadrature for $r'$ in the solution with the $\Lambda$-term}
\label{ap2}

We differentiate $r_{,_t}^2$ with respect to $r_i$:
\begin{equation}
\frac{d r_{,_t}^2}{dr_i}=2r_{,_t}r'_{,_t}=2r_{,_t}^2\frac{dr'}{dr}
\quad\Rightarrow\quad
\frac{dr'}{dr}=\frac{1}{2r_{,_t}^2}\cdot\frac{d r_{,_t}^2}{dr_i}
\label{p2-1}
\end{equation}
Substituting Eqn (\ref{5-2}) in (\ref{p2-1}) and differentiating
it with respect to $r_i$, we obtain the following equation for the
function ${y(r)\equiv r'}$:
\begin{equation}
\frac{dy}{dr}+P(r)y=Q(r) \, , \label{p2-2}
\end{equation}
Here, the functions ${P(r)\equiv P_1(r)+P_2(r)}$ and ${Q(r)\equiv
Q_1(r)+Q_2(r)}$ are given by (for convenience, $r_i$ is set to
unity below):
\begin{eqnarray}
P_1=\frac{-1}{2(r-1)}\, , \quad P_2=
\frac{1/(2r^2)-(r-1)/r^3-(a/2)(1-q^3/r^2)}{(1-r)/r^2+a(1+r+q^3/r)}
\, ,
\nonumber \\
Q_1=\frac{-1}{2(r-1)}\, , \quad Q_2=
\frac{1/(2r^2)+(r-1)/r^2+(a/2)(1-q^3/r)}{(1-r)/r^2
+a(1+r+q^3/r)}\, . \label{p2-3}
\end{eqnarray}
Equation (\ref{p2-2}) has a standard solution,
\begin{equation}
y(r)=\exp \left[ -\int\limits_{r_1}^r P(x)\, dx \right] \cdot
\left\{ \int\limits_{r_2}^r Q(r)\cdot\exp \left[
+\int\limits_{r_1}^r P(x)\, dx \right] \, dr +C_1 \right\}
\label{p2-4}
\end{equation}
The constants $C_1$, $r_1$ and $r_2$ are determined by the initial
conditions. The exponential can be rewritten as (the first term
with $P_1$ is integrated):
\begin{equation}
\exp \left[ \int\limits_{r_1}^r P(x)\, dx \right] =
\frac{C_2}{\sqrt{r-1}}\cdot\exp \left[ \int\limits_{1}^r P_2(x)\,
dx \right] \label{p2-5}
\end{equation}
After that, the part of integral (\ref{p2-4}) corresponding to
$Q_1$ can also be integrated (by parts), with the result
\begin{eqnarray}
y(r)=\sqrt{r-1}\cdot\exp\left[-\int\limits_1^r P_2(x)\, dx \right]
\cdot\left\{ \frac{1}{\sqrt{r-1}}\cdot \exp\left[+\int\limits_1^r
P_2(x)\, dx
 \right]\right|_{r_2}^r -
\nonumber \\
\left. -\int\limits_{r_2}^r \frac{P_2(r)}{\sqrt{r-1}}\cdot
\exp\left[+\int\limits_1^r P_2(x)\, dx\right] \,
dr+\int\limits_{r_2}^r \frac{Q_2(r)}{\sqrt{r-1}}\cdot
\exp\left[+\int\limits_1^r P_2(x)\, dx\right] \, dr +C_1\right\}
\label{p2-6}
\end{eqnarray}
We can then redefine the lower integration limit and determine the
constant $C_1$ from the initial conditions. The final expression
for the function ${y(r)=r'}$ is given in (\ref{rs2}).

$\quad $\\
\hrule

%\newpage

\end{document}